\documentclass[prd]{revtex4} 
 
\usepackage{amsmath} 
\usepackage{amssymb} 
\usepackage{graphicx} 
\usepackage{psfrag} 
 
\begin{document} 
 
\hspace*{14cm} ULB-TH/06-22 

\vskip 1cm

\title{MeV Right-handed Neutrinos and Dark Matter} 
 
\author{J.-M. Fr\`ere} 
\author{F.-S. Ling} 
\author{L. Lopez Honorez} 
\author{E. Nezri} 
\author{Q. Swillens} 
\author{G. Vertongen} 
\affiliation{Service de Physique Th\'eorique\\ 
Universit\'e Libre de Bruxelles\\ B-1050 Brussels, Belgium} 
 
\begin{abstract} 
We consider the possibility of having a MeV right-handed neutrino as 
a dark matter constituent. The initial reason for this study was the 
511 keV spectral line observed by the satellite experiment INTEGRAL: 
could it be due to an interaction between dark matter and baryons? 
Independently of this, we find a number of constraints on 
the assumed right-handed interactions. They arise in particular from 
the measurements by solar neutrino experiments. We come to the 
conclusion that such particles interactions are possible, and could 
reproduce the peculiar angular distribution, but not the rate of the 
INTEGRAL signal. However, we stress that solar neutrino experiments 
are susceptible to provide further constraints in the future. 
\end{abstract} 
 
 
\preprint{ULB-TH/06-22} 
 
\maketitle 
 
\section{Introduction} 
 
The initial motivation for this work comes from the INTEGRAL experiment. 
The existence of an excess of 511 keV photons~\cite{integral} has 
prompted a number of speculations, ranging from astrophysical~\cite{astro} 
to exotic~\cite{exotic} models. One of the 
characteristics of this radiation is its close association to the 
bulge area. A natural question arises then: could the radiation (or 
rather the production of positrons which induce it) be linked to an 
interaction of dark matter with baryonic matter which would also 
lead to a reduced emission region. 
 
The question of MeV neutrinos could be asked independently of this 
initial motivation and even be of some observational interest (Related works can be found in {\it e.g.}\cite{previous}.). In 
the following, we will consider neutrinos with suppressed weak 
interactions, essentially right-handed neutrinos interacting via 
$SU(2)_R$ gauge bosons. We will examine the stability, the 
cosmological abundance of such neutrinos and the constraints from 
neutrino experiments and finally discuss the 511 keV line signal. 
 
While such particles could account for the angular distribution, we 
find out that they cannot reproduce the rate of the INTEGRAL data. 
 
Nevertheless, we show that solar neutrino experiments can provide 
important constraints on such models. 
 
\section{Left-Right framework} 
 
Heavy right-handed neutrinos appear naturally in Left-Right (LR) symmetric models. We will 
assume the usual LR framework (see {\it e.g.} \cite{leftright}), in which the scalar sector is 
made of $\Delta_L(1,0,+2)$ and $\Delta_R(0,1,+2)$ which are respectively triplet under 
$SU(2)_L$ and $SU(2)_R$, and a bidoublet $\phi(\frac{1}{2},\frac{1}{2},0)$ under $SU(2)_L 
\otimes SU(2)_R \otimes U(1)_{B-L}$, in order to recover the Standard Model after symmetry 
breaking 
 
\begin{eqnarray}\label{groupe de jauge} 
SU(2)_L \otimes SU(2)_R \otimes U(1)_{B-L} \rightarrow SU(2)_L \otimes U(1)_Y  \rightarrow U(1)_Q. 
\end{eqnarray} 
 
At low temperatures, the physical gauge bosons are linear combinations of left and right-handed gauge bosons, 
 
\begin{displaymath} 
\left\{ 
\begin{array}{l} 
W_1 = \phantom{-}\cos\zeta ~W_L + \sin\zeta ~W_R \\ 
W_2 = - \sin\zeta ~W_L +  \cos\zeta ~W_R 
\end{array} \right. 
\qquad 
\left\{ 
\begin{array}{l} 
Z_1 = \phantom{-}\cos\chi ~Z_L + \sin\chi ~Z_R \\ 
Z_2 = - \sin\chi ~Z_L +  \cos\chi ~Z_R 
\end{array} \right. 
\end{displaymath} 
 
where the mixing angles $\zeta$ and $\chi$ can be expressed in terms of the vacuum expectation 
values (vev) of the scalar fields, and adequately suppressed to comply with experimental 
bounds. 
 
For the neutrinos, we invoke the see-saw mechanism to get light "left-handed", 
and heavy "right-handed" neutrinos. The mass eigenstates are actually mixtures of interaction states, 
 
\begin{displaymath} 
\left\{ 
\begin{array}{l} 
\nu_l = \phantom{-}\cos\theta~\nu_L + \sin\theta ~\nu_R \\ 
\nu_H = -\sin\theta ~\nu_L + \cos\theta~\nu_R 
\end{array} \right. 
\end{displaymath} 
with a mixing angle $\tan\theta \simeq (m_{\nu_l}/m_{\nu_H})^{1/2}$. 
It should be noted that it is possible in the LR framework to have a right-handed neutrino in the MeV range, 
while having right-handed gauge bosons masses in the TeV range. 
Compared to the case where heavy neutrino and right-handed gauge bosons masses are alike, 
this scenario does not seem as "natural" in the sense that it supposes tiny Yukawa couplings, 
but is still viable at this point. 
 
\section{Stability of the right-handed neutrinos} 
 
%\begin{figure}   
%\begin{center} 
%\psfrag{W}[c][c]{$W_{R,L}$} 
%\psfrag{g}[c][c]{$\gamma$} 
%\psfrag{n1}[c][c]{$\nu_R$} 
%\psfrag{n2}[c][c]{$\nu_L$} 
%\psfrag{l}[c][c]{$l$} 
%\includegraphics[height=.15\textheight]{plots/loop.ps} 
%\caption{A possible radiative decay of the right-handed neutrino.} 
%\label{decay} 
%\end{center} 
%\end{figure} 
For the MeV right-handed neutrino to be a dark matter candidate, its lifetime should be at least the age of the universe  $\tau_U \sim 1.4 \cdot 10^{10}$ years $\sim 4.5 \cdot 10^{17}$s. 
We first consider the decay channels $\nu_R \rightarrow \nu_L \nu_L  \bar{\nu}_L$ and $\nu_R \rightarrow e_R \bar{e}_L \nu_L$ due to LR mixing $\theta$ and $\zeta$. 
The decay widths at lowest order in $\theta, \zeta$ are 
\begin{equation} 
\Gamma_{\nu_R \rightarrow \nu_L \nu_L  \bar{\nu}_L}=\frac{\theta^2 G^2_F}{384 \pi^3}m^5_{\nu_H} \quad \quad; \quad \quad \Gamma_{\nu_R \rightarrow e_R \bar{e}_L \nu_L}=\frac{\zeta^2 G^2_F}{192 \pi^3}m^5_{\nu_H}. 
\end{equation} 
The requirement $\Gamma^{-1}>\tau_U$ implies very suppressed mixings, $\theta$ and $\zeta \leq 10^{-8}$. 
 
A radiative decay is also possible, namely $\nu_R \rightarrow \nu_L \gamma$. 
The decay width is proportional to the LR mixing $\zeta^2$, and to the mass square of the heaviest intermediate charged lepton that is coupled to $\nu_R$~\cite{decay}: 
\begin{equation} 
\Gamma_{\nu_R \rightarrow \nu_L \gamma} \simeq \frac{\alpha G_F^2}{2 \pi^4} \zeta^2 m^5_{\nu_H} 
\left( \frac{m_l^2}{m^2_{\nu_H}} \right)~. 
\end{equation} 
We can see that the constraint of stability from the radiative decay could amount to a further suppression of $\zeta$ by a factor $10^3$. 
When these requirements are fulfilled, we have $\nu_l \simeq \nu_L$ so that one of the three light neutrinos is almost massless $m_{\nu_l} \simeq 0$, $\nu_H \simeq \nu_R$ and  $W_1 \simeq W_L$, $W_2 \simeq W_R$. 
We note the charged gauge boson masses as $M_L$ and $M_R$ respectively, and their corresponding Fermi coupling constants for low-energy processes are 
\begin{equation} 
\frac{G_{L,R}}{\sqrt{2}}=\frac{g^2}{8M^2_{L,R}}~. 
\end{equation} 
 
\section{Cosmological abundance : low reheating temperature scenario}

As is well known, in the standard thermal freeze-out scenario, the so-called Lee-Weinberg bound forbids 
a neutrino with a mass in the MeV range and  a cross section of the order of weak interactions to be a cold dark matter relic. 
For a Majorana neutrino, its mass should be higher than $5$ GeV, in order to have 
a relic density $\Omega_\nu \leq 1$. A crude evaluation for the cold dark matter relic density is given by 
\begin{eqnarray}\label{omega approx} 
\Omega h^2 \approx \frac{3\times 10^{-27} \mbox{cm$^3$ s$^{-1}$}}{\langle \sigma v \rangle}~. 
\end{eqnarray} 
For right-handed neutrinos, the Lee-Weinberg limit is even further increased by a factor $G_R^2/G_L^2$. 
 
We are thus led to consider a low reheating temperature scenario for our dark matter 
candidate. Indeed, it has been noticed~\cite{riotto} that a decrease in 
the reheating temperature $T_{RH}$ can reduce the dark matter abundance. Usually, $T_{RH}$  is 
supposed to be high, but the only serious constraint is that the standard big-bang 
nucleosynthesis (BBN) scenario should not be spoiled, leading to $T_{RH} \geq 1$ MeV. 
 We do not consider the decay of the field responsible for the reheating  
and  we work with pre-BBN degrees of freedom to produce our dark matter neutrinos. We adopt an effective approach, in which neutrons and protons are produced during 
the reheating stage at the end of inflation,  right-handed neutrinos are  produced afterwards 
by interactions between nucleons and electrons.  In the following $T_{RH}$ will refer to the 
temperature at which dark matter production processes become effective; notice that it will 
take place well after inflation.  
%We are thus led to consider a low reheating temperature scenario instead. It has been 
%noticed~\cite{riotto} that a decrease in the reheating temperature $T_{RH}$ can reduce the 
%dark matter abundance. Usually, $T_{RH}$  is supposed to be high, but the only serious 
%constraint is that the standard big-bang nucleosynthesis (BBN) scenario should not be spoiled, 
%leading to $T_{RH} \geq 1$ MeV. In the following, we adopt an effective approach, in which 
%neutrons and protons are produced during the reheating stage at the end of inflation, while 
%right-handed neutrinos are only produced by interactions between nucleons and electrons. 
 
The right-handed neutrino relic abundance is obtained by solving numerically the Boltzmann 
equations after reheating  (see {\it e.g.}\cite{relic}), {\it i.e.} for $z \geq z_{RH}$, where 
$z=m_{\nu_R}/T$: 
\begin{eqnarray} 
\frac{H(m_{\nu_R})s(z)}{z}\frac{dY_{\nu}}{dz}& = & 
 -\gamma^{(pe\rightarrow 
  n\nu)}\left[\frac{Y_{\nu}}{Y^{eq}_{\nu}}\frac{Y_{n}}{Y^{eq}_{n}}-\frac{Y_p}{Y^{eq}_p}\right] 
-\gamma^{(ne\rightarrow 
  p\nu)}\left[\frac{Y_{\nu}}{Y^{eq}_{\nu}}\frac{Y_{p}}{Y^{eq}_{p}}-\frac{Y_n}{Y^{eq}_n}\right]  \nonumber\\ 
& & -\gamma^{(ee\rightarrow 
  \nu\nu)}\left[\frac{Y^2_{\nu}}{Y^{eq\ 2}_{\nu}}-1\right]  \nonumber 
\end{eqnarray} 
where $Y_{i}=n_i/s$ is the comoving density for the species $i$. 
The final right-handed neutrino relic density is $\Omega_{\nu_R}=n_{\nu}(T_{\gamma(2.7K)})m_{\nu_R}/\rho_c$. 
\begin{figure}[t!] 
\begin{center} 
\begin{tabular}{cc} 
\psfrag{mnu}[c][c]{\small $m_{\nu_R}~{\rm (MeV)}$} 
\psfrag{trh}[c][c]{\small $T_{RH}~{\rm (MeV)}$} 
\psfrag{fbt}[l][l]{\tiny From bottom to top~:} 
\psfrag{om}[c][c]{\scriptsize $~~~~~~~~~~\Omega_{\nu_R}=~0.01,~0.22,~1$}~ 
\psfrag{G1}[l][l]{\scriptsize $M_R/M_L=10$} 
\psfrag{G2}[l][l]{\scriptsize $M_R/M_L=20$} 
\includegraphics[width=0.42\textwidth]{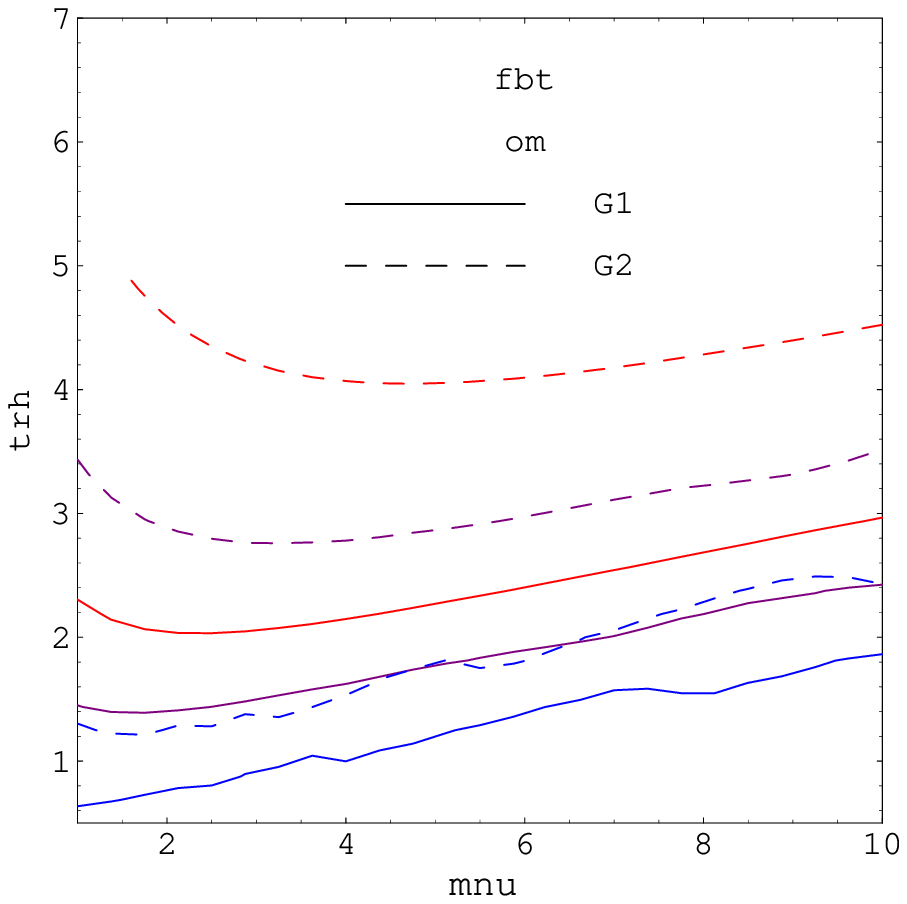}& 
\psfrag{T}[c][c]{\small ${\rm log_{10}}[T~(\rm MeV)]$} 
\psfrag{mnu}[c][c]{\scriptsize $m_{\nu_R}=4{\rm MeV}$} 
\psfrag{Omega}[c][c]{\small $\Omega_{\nu_R}$} 
\psfrag{G1}[l][l]{\scriptsize $M_R/M_L=20$} 
\psfrag{T1}[l][l]{\scriptsize $T_{RH}=2.84~{\rm MeV}$} 
\psfrag{G2}[l][l]{\scriptsize $M_R/M_L=10$} 
\psfrag{T2}[l][l]{\scriptsize $T_{RH}=1.65~{\rm MeV}$} 
\includegraphics[width=0.45\textwidth]{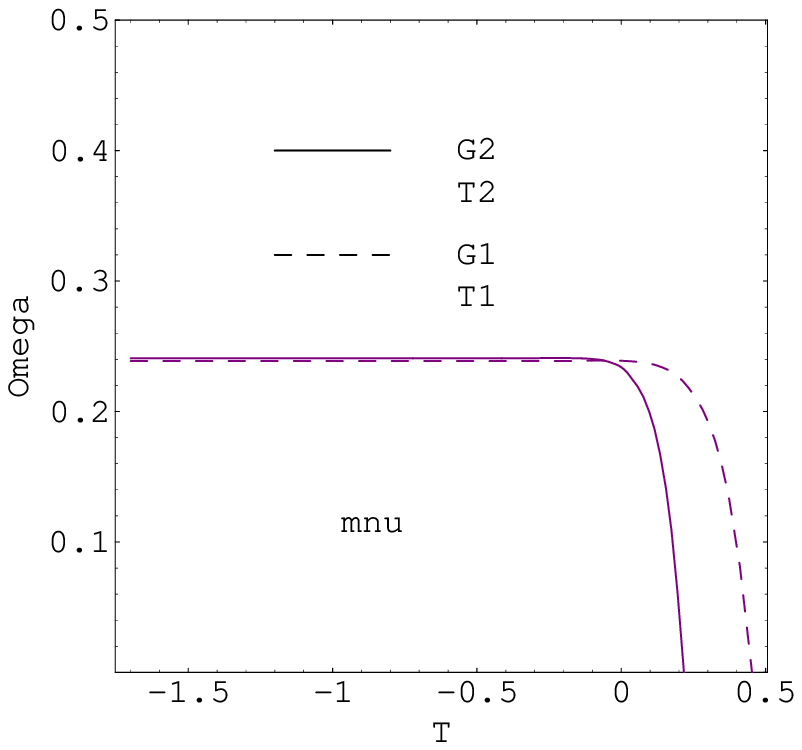}\\ 
(a) & (b) 
\end{tabular} 
\caption{a) Relic density of right-handed neutrino in the $(m_{\nu_R},T_{RH})$ plane. 
b) Solutions of the Boltzmann equation for $m_{\nu_R}=4~{\rm MeV}$ (in units of the critical density today).} 
\label{relic} 
\end{center} 
\end{figure} 
\noindent The results are shown on fig.~\ref{relic}. We see that for  $m_{\nu_R}~\!\sim~\!\mathcal{O}(\rm MeV)$, we can reach WMAP values ($\Omega_{\nu_R} \sim 0.2$) for 
 $T_{RH}\sim 1-10$ MeV which is quite small (see \cite{garcia} concerning  low scale inflation) and just above the BBN limit. 
For a given neutrino mass and reheating temperature, the relic density decreases when the charged gauge boson mass ratio $M_R/M_L$ is increased, because it is driven by the production cross section. 
Conversely, for a given neutrino mass and relic density, a higher reheating temperature is allowed when $M_R/M_L$ is increased. 
 
\section{Constraints from neutrino experiments} 
 
\begin{figure} 
\begin{center} 
\begin{tabular}{cc} 
\psfrag{nu}[c][c]{\small $\bar{\nu}_{R}$ } 
\psfrag{e}[c][c]{\small $e^+$ } 
\psfrag{W}[l][c]{\small $W_{R}$ } 
\psfrag{n}[c][c]{\small $p$ } 
\psfrag{p}[c][c]{\small $n$ } 
\includegraphics[width=0.22\textwidth]{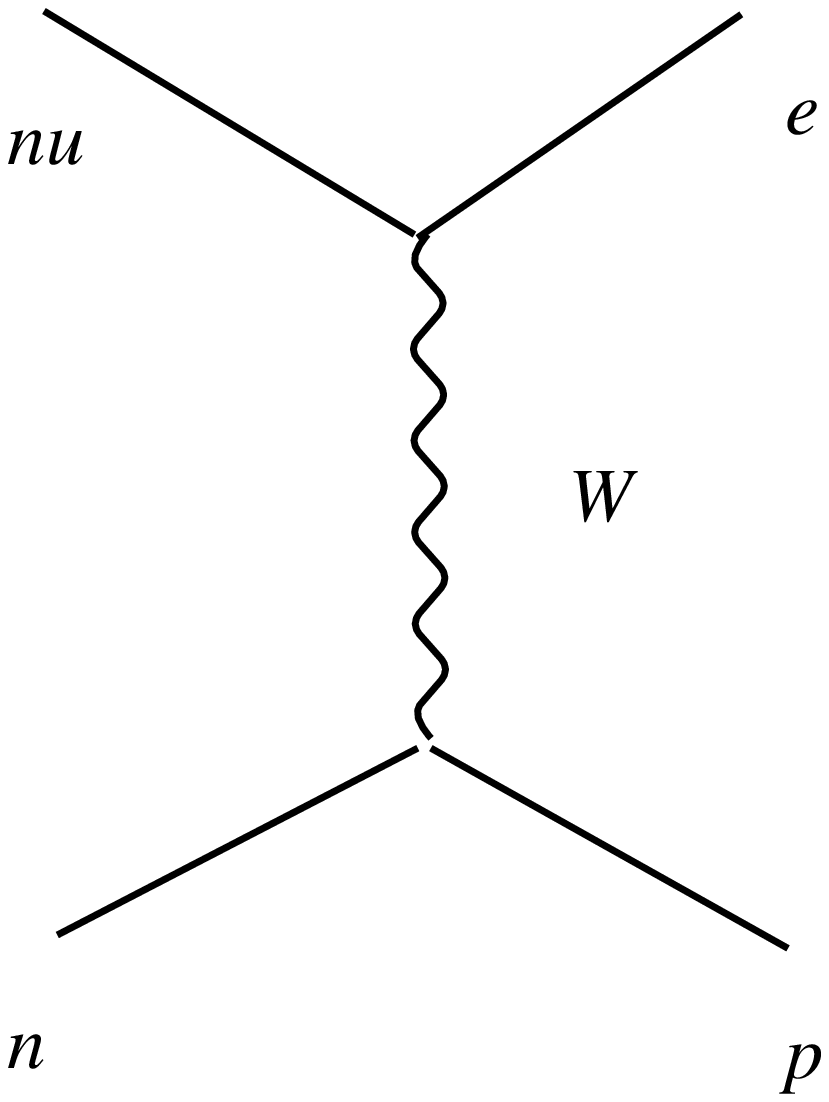}& 
\psfrag{nu}[c][c]{\small $\nu$ } 
\psfrag{e}[c][c]{\small $e^-$ } 
\psfrag{W}[l][c]{\small $W_{L,R}$ } 
\psfrag{n}[c][c]{\small $n$, $^{Z-1}\!A$} 
\psfrag{p}[c][c]{\small $p$, $^Z\!A$} 
\includegraphics[width=0.22\textwidth]{plots/WR.ps}\\ 
(a) & (b) 
\end{tabular} 
\caption{a) Process producing the positron for the 511 keV INTEGRAL signal. b) Left(Right) charged current driving $\nu_L +\, ^{Z-1}\!A\rightarrow e_L  +\,^{Z}\!A$ and $\nu_R +\,  ^{Z-1}\!A\rightarrow e_R  +\,^{Z}\!A$} 
\label{diag} 
\end{center} 
\end{figure} 
 
In our scenario, it turns out that the most stringent constraint on $M_R/M_L$ comes from solar neutrino chemical experiments. 
As can be seen in figure~\ref{diag}, the positron production reaction $\bar{\nu}_{R} + p \xrightarrow{W_R} \bar{e}_R +  n$ needed to explain the INTEGRAL signal, and 
the heavy neutrino capture reaction $\nu_{R} + n \ (^{Z-1}A) \xrightarrow{W_R} e_R +  p \ (^ZA)$ are closely related due to the Majorana nature of the neutrino. 
As the INTEGRAL diffuse gamma-ray data is compatible only with positron injection energies lower than 3 MeV~\cite{beyu}, 
we consider right-handed neutrinos with a mass lower than 5 MeV. 
Therefore, only the Gallium (Ga) experiments (Sage, Gallex, GNO) and Chlorine (Cl) experiments (Homestake) 
can be affected. The energy threshold in water Cerenkov experiments such as SuperKamiokande and SNO is higher than this value. 
The observed rates in Ga and Cl experiments are~\cite{nu01} 
\begin{eqnarray*} 
\left.\phi \sigma\right|_{\mbox{Ga}}=68.1\pm 3.85 \,~\mbox{SNU} \quad ; \quad \left.\phi \sigma\right|_{\mbox{Cl}}=2.56\pm 0.16 (stat) \pm 0.16(syst) ~\,\mbox{SNU}, 
\end{eqnarray*} 
with 1 SNU (Solar Neutrino Unit) = 10$^{-36}$ capture/atom/s. 
As these values are in good agreement with the results of Cerenkov water experiments (Kamiokande, SuperKamiokande and SNO), 
the possible right-handed contribution should fit inside the rate uncertainty. 
The total rate uncertainty is actually bigger than the experimental error quoted above, because one should add the uncertainties 
coming from the predicted rates with and without oscillation based on our knowledge of the solar model, the capture 
cross sections, and the results of SNO. 
It turns out that the total uncertainty at $1~\sigma$ amounts to $0.44$~SNU for the Cl experiment and $6.6$~SNU 
for Ga experiments. 
 
\begin{figure}[t!] 
\begin{center} 
\begin{tabular}{cc} 
\psfrag{M}[c][c]{\small $m_{\nu_R}$~(MeV)} 
\psfrag{MM}[c][c]{\small $M_R/M_L$} 
\psfrag{Ga}[c][c]{\small Ga} 
\psfrag{Cl}[c][c]{\small Cl} 
\includegraphics[height=.25\textheight]{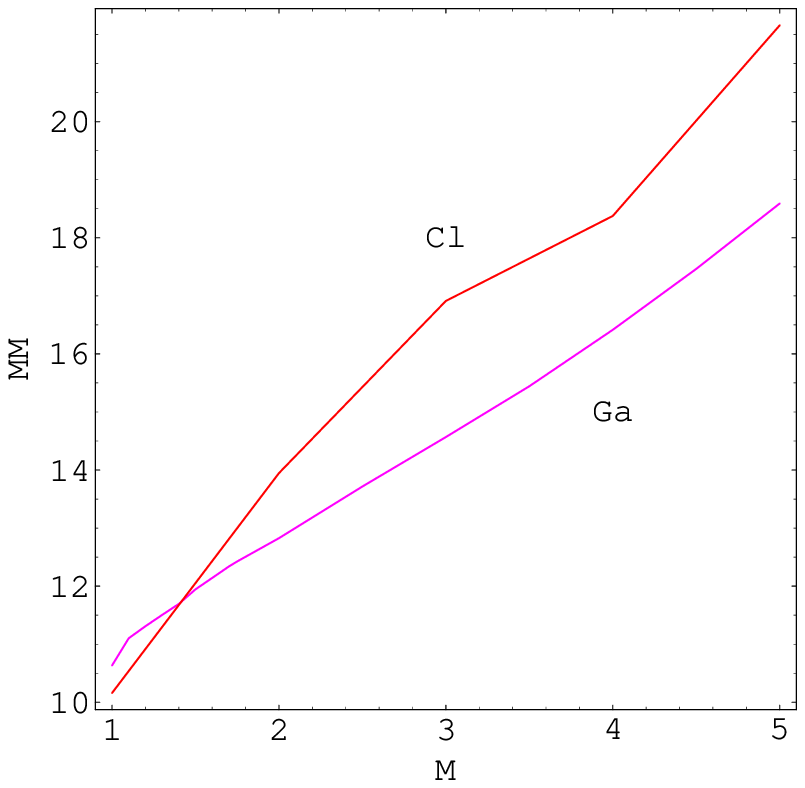}& 
\psfrag{M}[c][c]{\small $m_{\nu_R}$~(MeV) } 
\psfrag{sigma}[c][c]{\small $\sigma$ \footnotesize $(\nu_R + p \rightarrow e^+ + n)$~($\times 10^{-12}$ pb) } 
\includegraphics[height=.25\textheight]{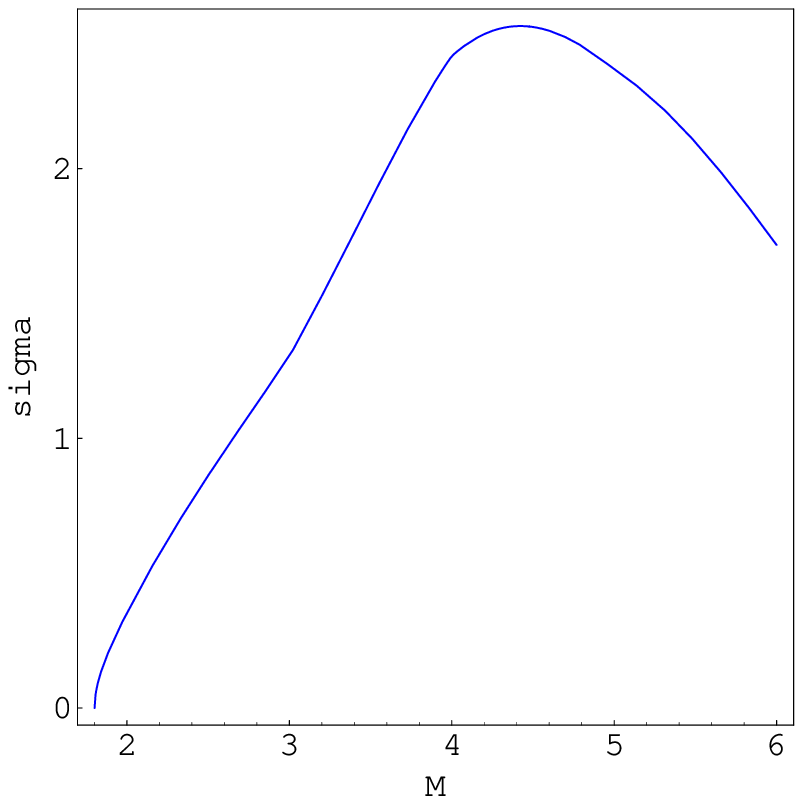}\\ 
(a) & (b) 
\end{tabular} 
\caption{a)$M_R/M_L$ lower bound from the solar neutrino chemical experiment data; b) 
Maximal $\nu_R$ - p cross section allowed by solar neutrino data} 
\label{nupcon} 
\end{center} 
\end{figure}

Assuming equality between left and right-handed couplings, 
we can relate the left and right-handed neutrino scattering cross sections 
$\sigma_{R,L} \equiv \sigma(\nu_{R,L} n \rightarrow e_{R,L} p)$, 
\begin{eqnarray} 
\sigma_R = \left(\frac{G_R}{G_L}\right)^2\frac{c}{|v_{\nu_R}|}~\sigma_L. 
\end{eqnarray} 
If the local $\nu_R$ density is set to $\rho_{dm}= 0.3$ GeV/cm$^3$, we have 
a $\nu_R$ flux $\phi_R = v_{\nu_R}\rho_{dm}/m_{\nu_R}$, so that 
\begin{eqnarray} 
\phi_R \sigma_R= 0.9 \,10^3\left(\frac{G_R}{G_L}\right)^2\left(\frac{{\rm MeV}}{m_{\nu_R}}\right) 
\left(\frac{\sigma_L}{10^{-10} {\rm pb}}\right) \, \mbox{SNU}. 
\end{eqnarray} 
As shown in fig.~\ref{nupcon}, the most constraining experiment is Chlorine. The values for $\sigma_L$ were taken from Bahcall~\cite{bahcall}. 
For $m_{\nu_R}\sim$ MeV we obtain $M_R/M_L \gtrsim 10-20$, which correspond to very suppressed cross sections, 
typically $\sigma_R \lesssim 10^{-12}$ pb. 
As a result, the $\nu_R$ positron production cross section is also limited in the same range. 
We find that it reaches a maximum of $2.4~10^{-12}$ pb for $m_{\nu_R} \simeq 4.5~{\rm MeV}$. 
It is interesting to note that we have a constraint on dark matter from neutrino experiments. 
This constraint is more stringent than direct accelerator limits, although less stringent than the much more speculative 
astrophysical bounds~\cite{pdg}. 
Our value relies on the assumption that the local dark matter density is around the 
usually quoted value of $0.3~ {\rm GeV/cm^3}$. If, for some reason, this density is well below this value, 
the constraint could fall behind other accelerator bounds on $M_R/M_L$. 
 
\section{Link with the 511 keV line signal} 
 
The gamma excess at 511 keV coming from the galactic bulge observed by INTEGRAL suggests electron-positron annihilations at rest. 
It is clear that the positrons have to be produced with a low energy in order to stay in the bulge. 
Different scenarios have been proposed to explain the positron production, but there is no definite answer. 
Here, we consider a neutrino-baryon ($\nu_R$-$B$) scattering through right-handed charged current interaction. 
The expected gamma ray flux in a solid angle $\Delta \Omega$ around the line of sight (l.o.s.) with direction $\vec{\psi}$ 
is given by 
\begin{equation} 
\phi(\vec {\psi},\Delta \Omega)=\frac{\sigma v}{2 \pi m_B m_{\nu_R}}\int_{\Delta\Omega} d \Omega \int_{l.o.s.} ds \ \rho_{DM} \rho_{B} 
\label{flux} 
\end{equation} 
 
\begin{figure} 
\begin{center} 
\psfrag{my2}[c][c]{\tiny 0.28} 
\psfrag{4.5}[c][c]{\tiny 0.63} 
\psfrag{my10}[c][c]{\tiny 1.4} 
\psfrag{my15}[c][c]{\tiny 2.14} 
\psfrag{pos}[c][l]{\tiny $>0$} 
\psfrag{neg}[c][l]{\tiny $<0$} 
\psfrag{psi}[t][b]{\small {\bf $\psi$ {\scriptsize (deg.)}}} 
\psfrag{J}[c][l][1][90]{\small {\bf Intensity} (ph $cm^{-2}s^{-1}rad^{-1}$)} 
\psfrag{rhoDM}[l][c]{\tiny  $\rho_{DM}= $NFW} 
\psfrag{rhobar1}[c][r]{ \tiny  $\rho_{bulge}= e^{-0.5r^2}$} 
\psfrag{rhobarDM}[tl][bl]{\tiny { $\rho_{DM}= $ Kravtsov}} 
 \includegraphics[height=.45\textwidth]{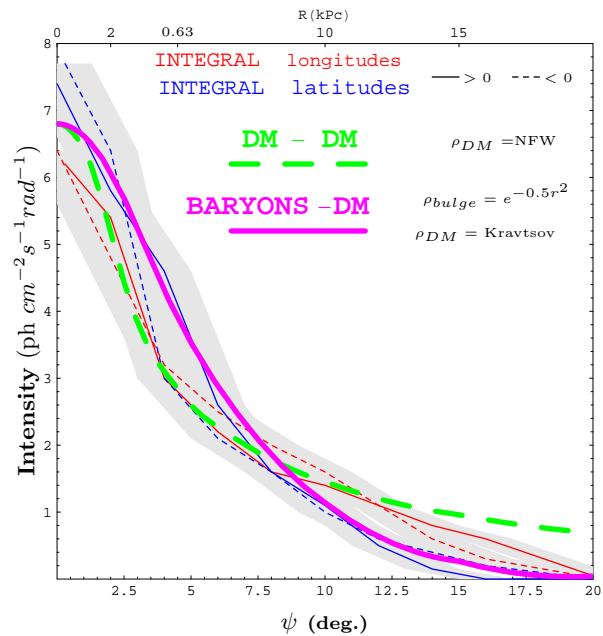} 
  \caption{Morphology of the estimated signal for dark matter-baryon interaction (thick magenta/dark) and for dark matter annihilation (thick green/dashed)  normalized to the observations vs. the INTEGRAL data (thin continuous and dashed lines + grey area).  $\psi$ is the angle between the line of sight and the galactic center direction. } 
\label{integral} 
\end{center} 
\end{figure}

Dark matter and baryon profiles are somewhat unknown in the central region of the galaxy. 
Considering the poor resolution of the INTEGRAL instrument ($\sim 3°$), the precise behavior near the galactic center 
cannot be probed. We have reported on fig.~\ref{integral} the INTEGRAL gamma-ray line flux data. 
The intensity profiles along positive and negative longitude and latitude were put together, assuming spherical symmetry, 
and broadened to the INTEGRAL resolution. 
 
For the DM halo, N-body simulations suggest the following effective parameterization 
\begin{equation} 
  \rho_{DM}(r)= \rho_{DM}^0 \frac{[1+(R_0/a)^{\alpha}]^{(\beta-\gamma)/\alpha}} 
  {[1+(r/a)^{\alpha}]^{(\beta-\gamma)/\alpha}}  \left(\frac{R_0}{r} \right)^{\gamma} , 
\label{eq:alphabetagamma} 
\end{equation} 
where $\gamma$ is quite unconstrained ($0\leq\gamma\leq 1.5$). 
 
For the baryonic profile of the bulge, we use the following function  \cite{freudenreich} 
\begin{equation} 
 \rho_B(r)= \rho^{0}_B \ e^{-0.5r^2}, 
\end{equation} 
where the value $\rho^{0}_B=30 \ {\rm GeV/cm^3}$ is derived from the galactic rotation 
curve~\cite{Sofue:2000jx} at the bulge scale. 
 
It turns out that the morphology of the expected signal,  when normalized  to the data, can reproduce the angular distribution of the INTEGRAL observations.  To 
illustrate this, fig.~\ref{integral} displays a  comparison between the
INTEGRAL data and the and the calculated gamma ray flux $\phi$ from positron
production (given by  eq.(\ref{flux})) in two different scenarios: 
\begin{itemize} 
\item light dark matter annihilation \cite{boehm} $\phi \propto \int \rho_{DM} \rho_{DM}$ with a NFW \cite{nfw} profile ($\gamma=1$ in eq.~(\ref{eq:alphabetagamma})), 
 
\item dark matter-baryon interaction $\phi \propto \int \rho_{B} \rho_{DM}$ with a Kravtsov  \cite{kra} profile ($\gamma=0.5$) for dark matter. 
Notice that the dark matter annihilation (through right-handed neutral currents) contribution 
is subdominant in this case \footnote{This contribution is suppressed by the choice of 
profiles and by the Majorana nature of $\nu_R$ in the annihilation cross section.}. 
 
\end{itemize} 
However, the magnitude of the measured flux (in a circle with 20 degrees radius) is \linebreak 
$(1.09\pm 0.04)\cdot 10^{-3} {\rm ~photons \ cm^{-2} ~s^{-1}}$, and cannot be reproduced in 
our model. Taking into account the neutrino experiment constraints on the cross section, we 
get a flux $\sim 10^{-13} {\rm ~photons \ cm^{-2}~s^{-1}}$, far too low to match the data.

\section{Conclusion} 
 
In a Left-Right framework, we studied the possibility of MeV right-handed Majorana neutrinos as dark matter candidate. 
The stability constraints can be fulfilled thanks to very suppressed left-right mixings for both bosons and neutrinos. 
The solar neutrino chemical experiments give the most stringent constraint on the right-handed interaction cross section, which appears to be very suppressed. 
Such a low cross section also excludes the standard freeze-out scenario. 
The WMAP relic abundance for the right-handed neutrino can still be achieved, at the cost of a very 
low reheating temperature scenario ($T_{RH}\sim \mathcal{O}({\rm 10 MeV})$). % This is quite an expensive price as it excludes any known baryogenesis mechanism. 
Finally, we have shown that the INTEGRAL 511 keV line signal morphology can be in good agreement with a dark matter-baryon interaction. 
In the specific model we considered, the right-handed neutrino is a viable dark matter candidate, but 
the cross section is too small to produce the right amount of photons. 
We point out that the careful comparison of solar neutrino experiments 
does and will provide stringent constraints on this and similar dark matter 
candidates. 
 
\section*{Acknowledgments} 
 
%FSL work is supported by a FNRS grant. EN 
We would like to thank N. Cosme and P. Vilain for helpful discussions. 
This work is supported by the FNRS (for FSL and EN), I.I.S.N. and the 
Belgian Federal Science Policy (return grant and IAP 5/27).

\end{document}